\documentclass[prl,amsfonts,twocolumn,nofootinbib,showpacs]{revtex4}
\usepackage{graphicx,epsfig}

\newcommand\G{\mbox{Gauss}}

\begin{document}

\title{Ellipsoidal Universe Can Solve The CMB Quadrupole Problem}

\author{L. Campanelli$^{1,2}$}
\email{campanelli@fe.infn.it}
\author{P. Cea$^{3,4}$}
\email{paolo.cea@ba.infn.it}
\author{L. Tedesco$^{3,4}$}
\email{luigi.tedesco@ba.infn.it}

\affiliation{$^1$INFN - Sezione di Ferrara, I-44100 Ferrara, Italy}
\affiliation{$^2$Dipartimento di Fisica, Universit\`{a} di Ferrara, I-44100 Ferrara, Italy}
\affiliation{$^3$INFN - Sezione di Bari, I-70126 Bari, Italy}
\affiliation{$^4$Dipartimento di Fisica, Universit\`{a} di Bari, I-70126 Bari, Italy}

\date{September, 2006}


\begin{abstract}
The recent three-year WMAP data have confirmed the anomaly
concerning the low quadrupole amplitude compared to the best-fit
$\Lambda$CDM prediction. We show that, allowing the large-scale
spatial geometry of our universe to be plane-symmetric with
eccentricity at decoupling or order $10^{-2}$, the quadrupole
amplitude can be drastically reduced without affecting higher
multipoles of the angular power spectrum of the temperature
anisotropy.
\end{abstract}


\maketitle


The latest results from the Wilkinson Microwave Anisotropy Probe
(WMAP)~\cite{3yearsWMAP} show that the cosmic microwave background
(CMB) anisotropy data are in remarkable agreement with the
simplest inflation model. At large scale, however, some anomalous
features has been reported. The most important discrepancy resides
in the low quadrupole moment, which signals an important
suppression of power at large scales. Note, however, that the
probability of quadrupole being low is not statistically
significant~\cite{3yearsWMAP}. Nevertheless, if this discrepancy
turns out to have a cosmological origin, then it could have far
reaching consequences for our understanding of the universe, and
in particular for the standard inflationary picture.
Indeed, it has been suggested that the low multipoles anomalies in
the CMB fluctuations may be a signal of a nontrivial cosmic
topology~\cite{Collins,Bunn,Topology}. For alternative solutions
to the quadrupole problem see Ref.~\cite{Quadrupole}, while for
other large scale anomalies in the angular distribution see
Ref.~\cite{Anomalies}.
\\
\indent
In this paper we show that the power suppression at large scales
can be accounted for if we relax the implicit assumption that the
large scale geometry is spherical. Indeed, if we assume that the
large-scale spatial geometry of our universe is plane-symmetric
with an eccentricity at decoupling of order $10^{-2}$, then we
find that the quadrupole amplitude can be drastically reduced with
respect to the value of the best-fit $\Lambda$CDM standard model
without affecting higher multipoles of the angular power spectrum
of the temperature anisotropy. This results is generic regardless
of the origin of  eccentricity.
\\
\indent
%
%
Let us begin by briefly discussing the standard analysis of the
temperature anisotropy~\cite{Dodelson}. First, the temperature
anisotropy is expanded in terms of spherical harmonics:
\begin{equation}
\label{DeltaT} \frac{\Delta T(\theta,\phi)}{\langle T \rangle} =
\sum_{l=1}^{\infty} \sum_{m=-l}^{l} a_{lm} Y_{lm}(\theta,\phi).
\end{equation}
%
%
After that, one introduces the power spectrum
\begin{equation}
\label{spectrum}
\frac{\Delta T_l}{\langle T \rangle} = \sqrt{
\frac{1}{2 \pi} \, \frac{l(l+1)}{2l+1} \sum_m |a_{lm}|^2},
\end{equation}
that fully determines all the properties of the CMB anisotropy. In
particular, the quadrupole anisotropy refers to the multipole
$l=2$:
\begin{equation}
\label{quadrupole-T} \mathcal{Q} \, \equiv \, \frac{\Delta
T_2}{\langle T \rangle} \, ,
\end{equation}
where $\langle T \rangle \simeq 2.73$K is the actual (average)
temperature of the CMB radiation. The quadrupole problem resides
in the fact that the observed quadrupole anisotropy
is~\cite{3yearsWMAP}:
\begin{equation}
\label{quad-obs} \left(\Delta T_2 \right)_{\rm obs}^{2} \simeq 211
\, \mu \mbox{K}^2,
\end{equation}
while the expected quadrupole anisotropy according the standard inflation is:
\begin{equation}
\label{quad-infl} \left(\Delta T_2 \right)_{\rm I}^2 \simeq 1252
\, \mu \mbox{K}^2.
\end{equation}
If we assume that the large-scale spatial geometry of our universe
is plane-symmetric with a small eccentricity, then following
Ref.~\cite{Bunn} we have that the observed CMB anisotropy map is a
linear superposition of two independent contributions
\begin{equation}
\label{quad-sum} \Delta T \; = \; \Delta T_{\rm A} + \, \Delta
T_{\rm I},
\end{equation}
where $\Delta T_{\rm A}$ represents the temperature fluctuations
due to the anisotropic space-time background, while $\Delta T_{\rm
I}$ is the standard isotropic fluctuation caused by the
inflation-produced gravitational potential at the last scattering
surface.  As a consequence, we may write:
\begin{equation}
\label{alm} a_{lm} = a_{lm}^{\rm A} + \, a^{\rm I}_{lm}.
\end{equation}
In this paper we will focus on the simpler case of plane-symmetric
space-time background. The most general plane-symmetric line
element~\cite{Taub} is:
\begin{equation}
\label{metric} ds^2 = dt^2 - a^2(t) (dx^2 + dy^2) - b^2(t) \,
dz^2,
\end{equation}
where we chose the $xy$-plane as the plane of symmetry. Here, the
scale factors $a$ and $b$ are functions of the cosmic time $t$
only. The most general energy-momentum tensor consistent with the
planar symmetry is of the form:
\begin{equation}
\label{tensor} T^{\mu}_{\;\; \nu} = \mbox{diag} \,
(\rho,-p_{\|},-p_{\|},-p_{\bot}),
\end{equation}
so that the Einstein's equations read:
\begin{eqnarray}
&& \left( \frac{\dot{a}}{a} \right)^{\!2} + 2 \,
\frac{\dot{a}}{a} \frac{\dot{b}}{b} = 8 \pi G \rho, \nonumber \\
\label{Einstein} && \frac{\ddot{a}}{a} + \frac{\ddot{b}}{b} +
\frac{\dot{a}}{a} \frac{\dot{b}}{b} = -8 \pi G p_{\|}, \\
&& 2 \, \frac{\ddot{a}}{a} + \left( \frac{\dot{a}}{a}
\right)^{\!2} = -8 \pi G p_{\bot} \nonumber,
\end{eqnarray}
where a dot indicates the derivative with respect to the cosmic
time. The total energy-momentum tensor $T^{\mu}_{\;\;\nu}$ can be
made up of two different components: an anisotropic contribution,
$(T_{\rm A})^{\mu}_{\;\;\nu} = \mbox{diag} \,
(\rho^{\rm A},-p_{\|}^{\rm A},-p_{\|}^{\rm A},-p_{\bot}^{\rm A})$,
which induces the planar symmetry --as, for example, a domain
wall, a cosmic string or a uniform magnetic field--, and an
isotropic contribution,
$(T_{\rm I})^{\mu}_{\;\;\nu} =
\mbox{diag} \, (\rho^{\rm I},-p^{\rm I},-p^{\rm I},-p^{\rm I})$,
such as vacuum energy, radiation, matter, or cosmological
constant. Exact solutions of Einstein's equations for different
kind of plane-symmetric plus isotropic components can be found in
Ref.~\cite{Berera}.
\\
In the following we will focus on the universe in the
matter-dominated era $p^{\rm I} = 0$ with  a plane-symmetric
component given by a uniform magnetic
field~\cite{Magnetic1,Magnetic2}. (Different physical models
giving rise to a plane-symmetric metric are discussed
in~\cite{Campanelli}.)
We will work in the limit of small eccentricity,
\begin{equation}
e = \sqrt{1 - (b/a)^2} \, ,
\end{equation}
and we normalize the scale factors such that $a(t_0) = b(t_0) = 1$
at the present time $t_0$. Moreover, we will suppose that, due to
the high conductivity of the primordial plasma, the magnetic field
is frozen into the plasma, so that it evolves as $B \propto
a^{-2}$ \cite{Magnetic1}.
The energy-momentum tensor for a uniform magnetic field can be
written as
$(T_{B})^{\mu}_{\;\;\nu} = \rho_B \, \mbox{diag} (1,-1,-1,1)$,
where $\rho_B = B^2/(8 \pi)$ is the magnetic energy density. Thus,
from the Einstein equations, we find that the eccentricity evolves
according to:
\begin{equation}
\label{evolution} \frac{d (e \dot{e})}{dt} + 3 H (e \dot{e}) = 16
\pi G \rho_B,
\end{equation}
where $H = \dot{a}/a$. In the matter-dominated era and at the
lowest order in $e$, we have $a \propto t^{2/3}$ and then $H =
2/(3t)$. The solution of Eq.~(\ref{evolution}), with the condition
$e(t_0) = 0$, is
$e^2 = 8 \Omega_B^{(0)} ( 1 - 3 a^{-1} + 2 a^{-3/2})$,
where $\Omega_B^{(0)} = \rho_B(t_0)/\rho_{\rm cr}^{(0)}$, and
$\rho_{\rm cr}^{(0)} = 3 H_0^2/8 \pi G$ is the actual critical
energy density. At the decoupling, $t=t_{\rm dec}$, we have
$e_{\rm dec}^2 \simeq 16 \, \Omega_B^{(0)} z_{\rm dec}^{3/2}$,
where $e_{\rm dec} = e(t_{\rm dec})$ and $z_{\rm dec} \simeq 1088$
is the red-shift at decoupling \cite{1yearWMAP}. As a consequence,
we get:
\begin{equation}
\label{eccentricity2} e_{\rm dec} \simeq 10^{-2} h^{-1}
\frac{B_0}{10^{-8} \G} \: ,
\end{equation}
where $B_0 = B(t_0)$ and $h \simeq 0.72$ \cite{1yearWMAP} is the
little-$h$ constant.
\\
We are interested in the distortion of the CMB radiation in a
universe with planar symmetry described by the metric
(\ref{metric}). As before, we will work in the small-eccentricity
approximation. From the null geodesic equation, we get that a
photon emitted at the last scattering surface having energy
$E_{\rm dec}$ reaches the observer with an energy equal to
$E_0(\widehat{n}) = \langle E_0 \rangle (1 - e_{\rm dec}^2 n_3^2/2)$,
where $\langle E_0 \rangle \equiv E_{\rm dec}/(1+z_{\rm dec})$,
and $\widehat{n} = (n_1,n_2,n_3)$ are the direction cosines of the
null geodesic in the symmetric (Robertson-Walker) metric.
\\
It is worth mentioning that the above result applies to the case
of a magnetic field directed along the $z$-axis. We may, however,
easily generalize this result to the case of a magnetic field
directed along an arbitrary direction in a coordinate system
$(x_{\rm g},y_{\rm g},z_{\rm g})$ in which the $x_{\rm g} y_{\rm
g}$-plane is, indeed, the galactic plane. To this end, we perform
a rotation $\mathcal{R} = \mathcal{R}_{x}(\vartheta) \,
\mathcal{R}_{z}(\varphi + \pi/2)$ of the coordinate system
$(x,y,z)$, where $\mathcal{R}_{z}(\varphi + \pi/2)$ and
$\mathcal{R}_{x}(\vartheta)$ are rotations of angles $\varphi +
\pi/2$ and $\vartheta$ about the $z$- and $x$-axis, respectively.
In the new coordinate system the magnetic field is directed along
the direction defined by the polar angles $(\vartheta, \varphi)$.
Therefore, the temperature anisotropy in this new reference system
is:
\begin{equation}
\label{DeltaTA} \frac{\Delta T_{\rm A}}{\langle T \rangle} \equiv
\frac{\langle E_0 \rangle - E_0(n_{\rm A})}{\langle E_0 \rangle} =
\frac{1}{2} \, e_{\rm dec}^2 n_{\rm A}^2,
\end{equation}
where $n_{\rm A} \equiv (\mathcal{R} \, \widehat{n})_3$ is equal to
\begin{equation}
\label{nA} n_{\rm A}(\theta,\phi) = \cos \theta \cos \vartheta -
\sin \theta \sin \vartheta \cos(\phi - \varphi).
\end{equation}

Alternatively, when the ellipticity is small, Eq.~(\ref{metric})
may be written in a more standard form:
\begin{equation}
\label{metric2} ds^2 = dt^2 - a^2(t) (\delta_{ij} + h_{ij}) \,
dx^i dx^j,
\end{equation}
where $h_{ij}$ is a metric perturbation which takes on the form
$h_{ij} = -e^2 \delta_{i3} \delta_{j3}$.
The null geodesic equation in a perturbed
Friedman-Robertson-Walker metric gives the temperature anisotropy
(Sachs-Wolfe effect):
\begin{equation}
\label{anis2} \frac{\Delta T}{\langle T \rangle}  =  - \frac{1}{2}
\int_{t_0}^{t_{\rm dec}} \! dt \; \frac{\partial h_{ij}}{\partial
t} \; n^i n^j ,
\end{equation}
where $n^i$ are the directional cosines. Using $e(t_0)=0$, from
Eqs.~(\ref{metric2}) and (\ref{anis2}) one gets
$\Delta T/\langle T \rangle = (1/2)\, e^2_{\rm dec}n_3^2$,
which indeed agrees with the above result.

Before proceeding further, we note that the result (\ref{DeltaTA})
is subject to a simple geometrical interpretation. Indeed, let the
surface of last scattering be an ellipsoid with semi-axes $a_{\rm
dec}$, $b_{\rm dec}$, and eccentricity $e_{\rm dec} = \sqrt{1
-(b_{\rm dec}/a_{\rm dec})^2}$ (see Fig.~1). Then, a photon
starting from the last scattering surface at the emitter point E
with spherical coordinates $(r,\theta,\phi)$ reaches the observer
O with an energy proportional to $r$, $E_0(r) = E_{\rm dec} \,
r/a_{\rm dec}$. Therefore, taking into account the equation of an
ellipsoid whose $b$-axis is directed along the direction defined
by the polar angles $(\vartheta, \varphi)$,
\begin{equation}
\frac{a_{\rm dec}^2}{r^2} = 1 + \frac{e_{\rm dec}^2}{1-e_{\rm
dec}^2} \: n_{\rm A}^2(\theta,\phi),
\end{equation}
we recover, for small eccentricity, the previous results.


\begin{figure}[t]
\begin{center}
\includegraphics[clip,width=0.35\textwidth]{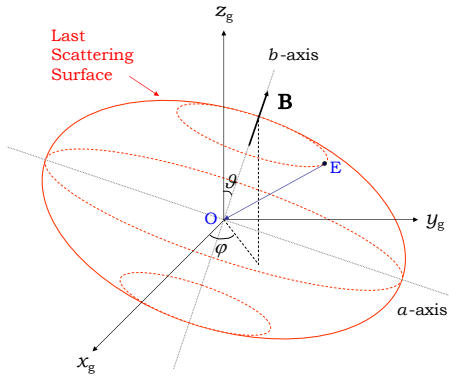}
\caption{Geometrical interpretation of the CMB anisotropy in an
``ellipsoidal'' universe. A photon emitted at the point E of the
ellipsoidal surface of last scattering reaches the observer O with
an energy proportional to the distance between emitter and
observer. This causes a quadrupole anisotropy in the CMB
radiation.}
\end{center}
\end{figure}


It is easy to see  from Eq.~(\ref{DeltaTA}) that only the
quadrupole terms ($l=2$) are different from zero:
\begin{eqnarray}
\label{almA} && a_{20}^{\rm A} = \frac{\sqrt{\pi}}{6\sqrt{5}} \,
             [1 + 3\cos(2 \vartheta) ] \, e_{\rm dec}^2 \, , \nonumber \\
             && a_{21}^{\rm A} = -(a_{2,-1}^{\rm A})^{*} =
             -\sqrt{\frac{\pi}{30}} \;
             e^{-i \varphi}  \sin(2\vartheta) \, e_{\rm dec}^2 \, , \nonumber \\
             && a_{22}^{\rm A} = (a_{2,-2}^{\rm A})^{*} =
             \sqrt{\frac{\pi}{30}} \; e^{-2 i \varphi} \sin^2\!\vartheta \,
             e_{\rm dec}^2 \, .
\end{eqnarray}
Consequently, the quadrupole anisotropy is
\begin{equation}
\mathcal{Q}_{\rm A} = \frac{2}{5 \sqrt{3}} \; e_{\rm dec}^2 \, .
\end{equation}
Since the temperature anisotropy is a real function, we have
$a_{l,-m} = (-1)^m (a_{l,-m})^*$. Observing that $a_{l,-m}^{\rm A}
= (-1)^m (a_{l,-m}^{\rm A})^*$ [see Eq.~(\ref{almA})], we get
$a^{\rm I}_{l,-m} = (-1)^m (a^{\rm I}_{l,-m})^*$.
Moreover, because the standard inflation-produced temperature
fluctuations are statistically isotropic, we will make the
reasonable assumption that the $a^{\rm I}_{2m}$ coefficients are
equals up to a phase factor. Therefore, we can write:
\begin{eqnarray}
\label{almI}
&& a^{\rm I}_{20} = \sqrt{\frac{\pi}{3}} \; e^{i \phi_1} \mathcal{Q}_{\rm I}, \nonumber \\
&& a^{\rm I}_{21} =  - (a^{\rm I}_{2,-1})^{*} =
                    \sqrt{\frac{\pi}{3}} \; e^{i \phi_2} \mathcal{Q}_{\rm I}, \\
&& a^{\rm I}_{22} =  (a^{\rm I}_{2,-2})^{*} =
                     \sqrt{\frac{\pi}{3}} \; e^{i \phi_3} \mathcal{Q}_{\rm I}, \nonumber
\end{eqnarray}
where $0 \leq \phi_i \leq 2 \pi$ are arbitrary phases. Taking into
account Eqs.~(\ref{almA}) and (\ref{almI}), and
Eqs.~(\ref{DeltaT}), ~(\ref{spectrum}) and~(\ref{alm}), we get for
the total quadrupole:
\begin{equation}
\label{quadrupole} \mathcal{Q}^2 = \mathcal{Q}_{\rm A}^2 +
\mathcal{Q}_{\rm I}^2 - 2f \mathcal{Q}_{\rm A} \mathcal{Q}_{\rm
I},
\end{equation}
where
\begin{eqnarray}
\label{f}  f(\vartheta, \varphi \, ; \phi_1, \phi_2, \phi_3) & = &
\frac{1}{4 \sqrt{5}} \, \{
           2 \sqrt{6} \, \left[-\sin\vartheta \cos(2\varphi +
           \phi_3) \right.
\nonumber \\
           & + &
           \left. 2 \cos\vartheta \cos(\varphi + \phi_2) \right] \nonumber \\
           & - & [1 + 3 \cos(2 \vartheta) ] \cos\phi_1 \}.
\end{eqnarray}
Looking at Eq.~(\ref{quadrupole}) we see that, if the space-time
background is not isotropic, the quadrupole anisotropy can become
smaller than the one expected in the standard picture of the
$\Lambda$CDM (isotropic-) cosmological model of temperature
fluctuations. We may fix the direction of the magnetic field and
the eccentricity by minimizing the total quadrupole anisotropy.
Let  $\bar{e}_{\rm dec}$, $\bar{\vartheta}$, and $\bar{\varphi}$
be the values which minimize $\mathcal{Q}^2$, and $\bar{f}$ the
expression $f(\bar{\vartheta}, \bar{\varphi} \, ; \phi_1, \phi_2,
\phi_3)$, then we get:
\begin{equation}
\label{quadrupole-ecc-min} \mathcal{Q}^2 = (1 - \bar{f}^2)
\mathcal{Q}_{\rm I}^2, \;\;\;\; \bar{e}_{\rm dec}^2 =
\frac{5\sqrt{3}}{2} \: \bar{f} \, \mathcal{Q}_{\rm I}.
\end{equation}
It is straightforward to show that $\bar{f}$ is a strictly
positive function, such that, for every $\bar{\vartheta}$,
$\bar{\varphi}$, $\phi_1$, $\phi_2$, and $\phi_3$, it results:
\begin{equation}
\label{fbar} \frac{1}{\sqrt{5}} \leq \bar{f} \leq \frac{\sqrt{39 +
6 \sqrt{6}} + \sqrt{6} - 1}{4 \sqrt{5}} \: .
\end{equation}
Taking $\left(\Delta T_2 \right)^2_{I} = 1252 \, \mu \mbox{K}^2$,
we have
\begin{equation}
\label{QLimit} 46.2 \, \mu \mbox{K}^2 \lesssim \left(\Delta T_2
\right)^{2} \lesssim 1001.6 \, \mu \mbox{K}^2,
\end{equation}
\begin{equation}
\label{eLimit} 0.50 \times 10^{-2} \lesssim \bar{e}_{\rm dec}
\lesssim 0.74 \times 10^{-2}.
\end{equation}
From the above equations we see that the inflation-produced
quadrupole anisotropy embedded in a plane-symmetric universe with
eccentricity of order $10^{-2}$ can be brought into agreement with
observations (see Fig.~2). In particular, from
Eq.~(\ref{eccentricity2}) it follows  that the eccentricity
$\bar{e}_{\rm dec} \simeq (0.5 \div 0.7) \times 10^{-2}$ is
produced by cosmic magnetic fields $B_0 \simeq (4 \div 5) \times
10^{-9} \G$ (these values for $B_0$ are compatible with the
constraints on cosmic magnetic fields derived in
Ref.~\cite{Barrow}).


\begin{figure}
\begin{center}
\includegraphics[clip,width=0.40\textwidth]{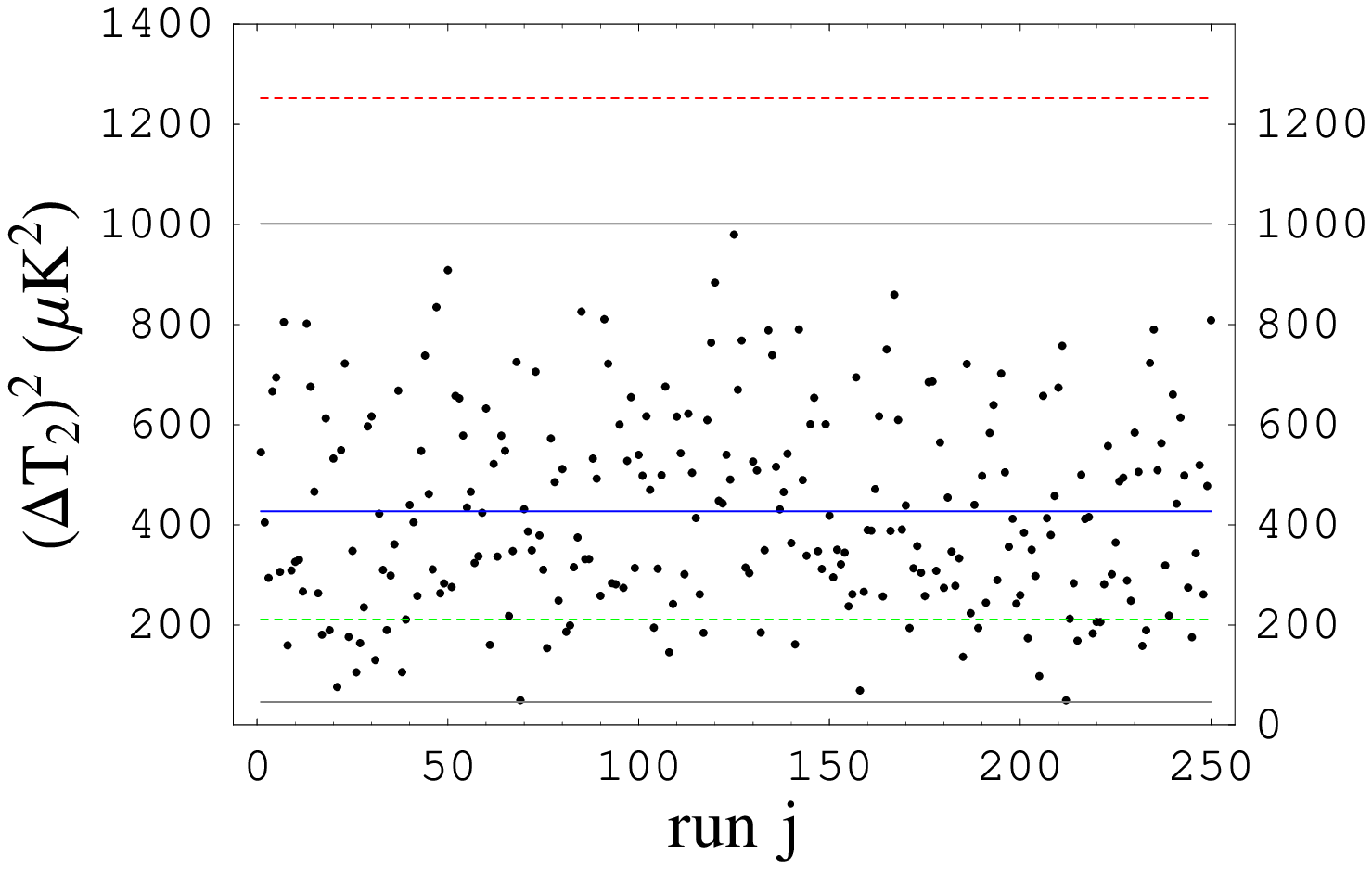}
\includegraphics[clip,width=0.40\textwidth]{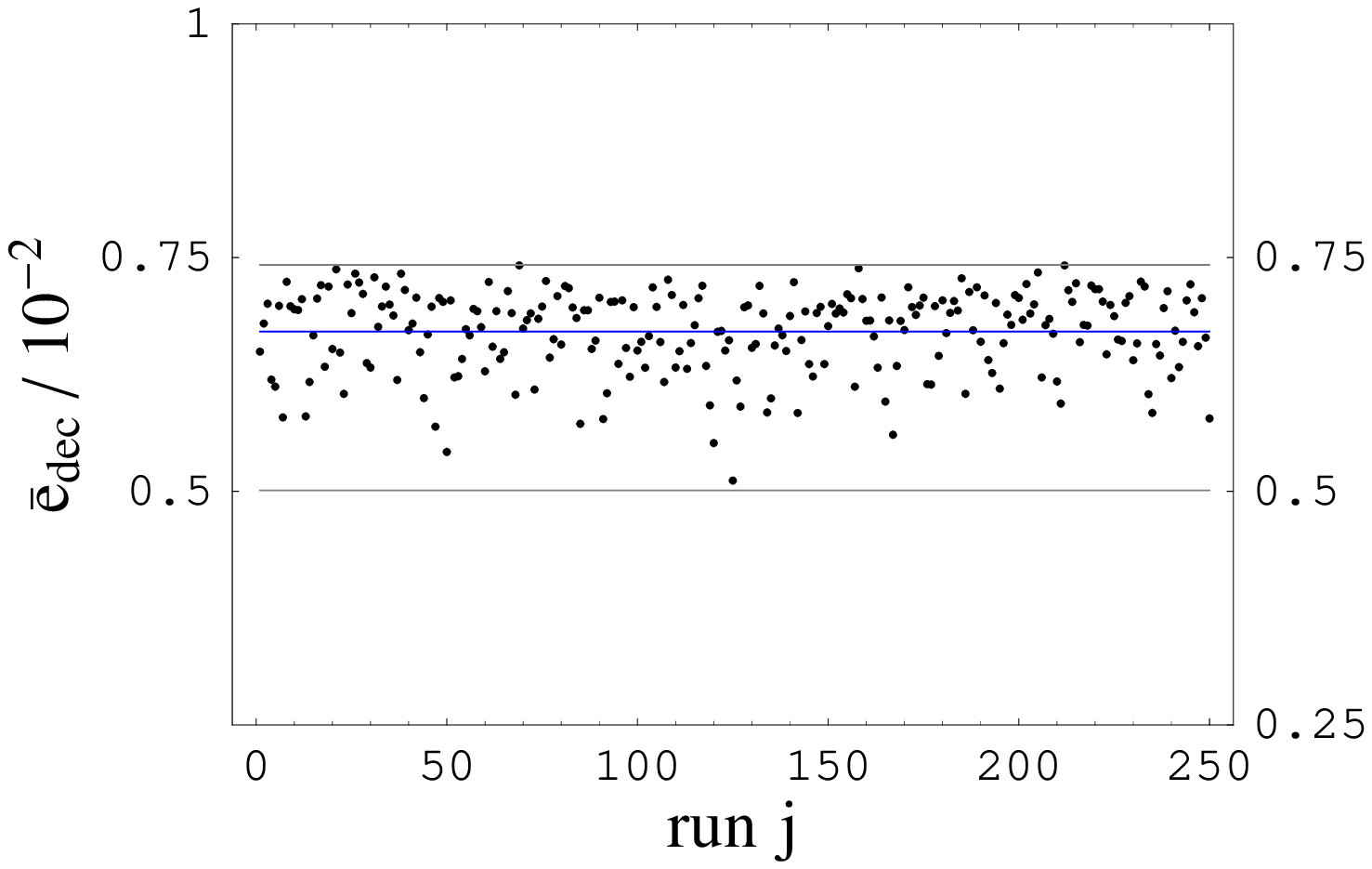}
\caption{Results from our numerical simulation with $N=250$ runs.
Continuous grey lines in the upper and lower panel refer to
Eqs.~(\ref{QLimit}) and (\ref{eLimit}), respectively.
{\it Upper panel}. The blue continuous line is the mean value
Eq.~(\ref{meanT}), the green dashed line is Eq.~(\ref{quad-obs}),
while the red dashed line is Eq.~(\ref{quad-infl}).
{\it Lower panel}. The blue continuous line is the mean value
Eq.~(\ref{meane}).}
\end{center}
\end{figure}


Obviously, the arbitrary phases $\phi_i$ are not directly
measurable. Therefore, we can treat them as stochastic variables.
As a consequence, the quadrupole $\mathcal{Q}$,
Eq.~(\ref{quadrupole}), can be considered as a distribution
function which depends on the parameters $e_{\rm dec}$,
$\vartheta$, $\varphi$ and the stochastic variables $\phi_i$. To
determine the distribution of the quadrupole, we have performed
numerical simulations keeping the variables $\phi_i$ random in the
interval  $[0,2 \pi]$. In Fig.~2 we show the result of our
numerical simulations with $N=250$ runs. In each run, according to
our previous discussion, $e_{\rm dec}$, $\vartheta$, $\varphi$ are
determined by minimizing the total quadrupole. We find for the
arithmetic averages:
\begin{equation}
\label{meanT} [(\Delta T_2)^2]_{\rm mean} \simeq 427.3 \, \mu
\mbox{K}^2,
\end{equation}
\begin{equation}
\label{meane} (\bar{e}_{\rm dec})_{\rm mean} \simeq 0.67 \times
10^{-2}.
\end{equation}
Equation~(\ref{meanT}) shows that, even if we take care only of
the intrinsic uncertainty measured by the cosmic variance:
\begin{equation}
\label{cosmic} \sigma_{\rm cosmic} \equiv \sqrt{2/5} \: [(\Delta
T_2)^2]_{\rm mean} \simeq 270.3 \, \mu \mbox{K}^2,
\end{equation}
then our mean value is in agreement with the observed quadrupole
anisotropy Eq.~(\ref{quad-obs}). \\
\indent
In conclusion, we have shown that allowing a small eccentricity at
decoupling $e_{\rm dec} \sim 10^{-2}$ in the large-scale spatial
geometry of our universe, it results in a drastic reduction in the
quadrupole anisotropy without affecting higher multipoles of the
angular power spectrum of the temperature anisotropy. Moreover, we
have seen that such a small eccentricity could be generated by a
uniform cosmic magnetic field whose actual strength, $B_0 \sim
10^{-9} \, \G$, is of the correct order of magnitude to account
for the observed magnetic fields in galaxies and galaxy clusters.



\begin{acknowledgments}
L. C. thanks M. Giannotti and F. L. Villante for helpful
discussions. This work is supported in part by the Italian INFN
and MIUR through the ``Astroparticle Physics'' research project.
\end{acknowledgments}


\end{document}